\documentclass[a4paper,reqno,12pt,final]{amsart}
\usepackage{amsmath,amssymb,eucal,bbold}
\usepackage[T1]{fontenc}
\usepackage[latin1]{inputenc}
\usepackage{array}
\usepackage{epsfig}
\newcommand{\into}{\hookrightarrow}
\newcommand{\Aut}{\mathrm{Aut}}
\newcommand{\Inn}{\mathrm{Inn}}
\newcommand{\Out}{\mathrm{Out}}
\newcommand{\eU}{\EuScript{U}}
\newcommand{\eV}{\EuScript{V}}
\newcommand{\RR}{\mathbb R}
\newcommand{\ZZ}{\mathbb Z}
\newcommand{\SU}{\mathrm{SU}}
\newcommand{\U}{\mathrm{U}}
\begin{document}

\title[]{D-brane charge, flux quantisation and relative (co)homology}
\author[Figueroa-O'Farrill]{José M Figueroa-O'Farrill}
\address{\begin{flushright}
    Department of Mathematics and Statistics\\
    The University of Edinburgh\\
    Edinburgh EH9 3JZ\\
    Scotland\\
  \end{flushright}}
\email{jmf@maths.ed.ac.uk}
\author[Stanciu]{Sonia Stanciu}
\address[]{\begin{flushright}Spinoza Institute\\
Utrecht University\\
Leuvenlaan 4\\
3508 TD Utrecht\\
The Netherlands\\
\end{flushright}}
\email{s.stanciu@phys.uu.nl} \thanks{Edinburgh MS-00-010, Spin
  2000/19, \texttt{arXiv:hep-th/0008038}}
\begin{abstract}
  We reconsider the problem of $\U(1)$ flux and D0-charge for D-branes
  in the WZW model and investigate the relationship between the
  different definitions that have been proposed recently.  We identify
  the D0-charge as a particular reduction of a class in the relative
  cohomology of the group modulo the D-submanifold.  We investigate
  under which conditions this class is equivalent to the first Chern
  class of a line bundle on the D-submanifold and we find that in
  general there is an obstruction given by the cohomology class of the
  NS $3$-form.  Therefore we conclude that for topologically
  nontrivial $B$-fields, there is strictly speaking no $\U(1)$ gauge
  field on the D-submanifold.  Nevertheless the ambiguity in the flux
  is not detected by the D0-charge.  This has a natural interpretation
  in terms of gerbes.
\end{abstract}
\maketitle

\section{Introduction}

D-branes in Lie groups have received a great deal of attention
recently.  They provide an ideal laboratory for the study of D-branes
in nontrivial string backgrounds, as they are amenable to both
microscopic analysis via the algebraic methods of boundary conformal
field theory and to the more standard field theoretic techniques based
on sigma models and the path integral.  Moreover it is precisely in
the combination of these two rather different approaches that many of
the recent developments have taken place.

One of these recent developments concerns the problem of stability of
D-branes and the definition and quantisation of their D0-charge
\cite{BDS,Pw,Ta,KKZ,AMM,Marolf1,Q0,AS2,Pelc,Marolf2}, about which
several somewhat different points of view seem to emerge.  Our
motivation in this paper is to further explore these points of view.

Throughout this paper we will refer to the submanifolds on which
D-branes can wrap as D-submanifolds.  In the case of the WZW model,
the D-submanifolds in question are described by (twisted, shifted)
conjugacy classes \cite{AS,SAdS3,FFFS,SDnotes}.  In general these
submanifolds are not minimal (although see \cite{FSS3} for an example
of a totally geodesic twisted conjugacy class in $\SU(2)\times\SU(2)$)
and hence they are not stabilised gravitationally.  This is hardly
surprising as the metric is not the only field in the background:
there is also a $B$-field, or more precisely a closed $3$-form $H$
with integral periods.

The stability of a D-brane wrapping such submanifolds should therefore
be related to this integrality condition and should manifest itself in
the quantisation of the allowed D-submanifolds. There seem to be at
least two mechanisms through which this may be explained.  One of them
\cite{AS,Gaw,Pw,Q0} uses an argument which can be understood as the
vanishing of the global worldsheet anomaly \cite{FW} in the lagrangian
description of the boundary WZW model.  Alternatively, it was argued
in \cite{BDS} that the stability of these D-branes is a result of the
quantisation of the flux of a $\U(1)$ gauge field on the D-brane.

As far as the definition of the D0-charge is concerned there seem to
be a number of candidates.  As we will recall below, the data
describing a D-submanifold in a Lie group is a pair $(Q,\omega)$ where
$Q$ is, in the simplest case, a conjugacy class and $\omega$ is a
two-form on $Q$ such that $d\omega = H$ there.  According to one view
\cite{BDS,Pw}, the D0-charge of the D-brane is defined by the flux of
the gauge invariant two-form field $\omega$ on the D-submanifold.  If
one computes explicitly the spectrum of this D0-charge for the
possible D2-branes in $\SU(2)$, one obtains that these charges are not
quantised; however their values agree with the RR charges as
obtained from the boundary state approach.  This is the charge
recently identified in \cite{Marolf1} as the brane source charge.

A second candidate definition of the D0-charge is obtained
\cite{Ta,AMM} by adding to the brane source charge a bulk
contribution, such that the resulting D0-charge agrees with the flux
of a $\U(1)$ gauge field defined on the D-submanifold.  According to
this view, the quantisation of the D0-charge is a consequence of the
quantisation of the flux of this $\U(1)$ field.  In the nomenclature
of \cite{Marolf1}, this charge can be identified either with the Page
charge \cite{AMM} or with the Maxwell charge \cite{Ta}, which in this
case coincide.  This definition of the D0-charge requires a
trivialisation of the NS $3$-form.

An alternative definition of the D0-charge of such a D-brane was
proposed in \cite{Q0} by one of the authors, where the quantisation of
the D0-charge is seen as a consequence of the vanishing of the global
worldsheet anomaly for the boundary WZW model.  This third definition
is essentially a Page charge, and one of the aims of this paper is to
compare this charge with the one in \cite{Ta,AMM}.

Let $Q$ be a D-submanifold in a compact simple Lie group $G$ and let
$k$ be the level of the WZW model.  Let $H$ be the bi-invariant
$3$-form in $G$ and let $\omega$ be a $2$-form on $Q$ such that
$H=d\omega$ there.  We will show that the charge in \cite{Q0} can be
identified with the reduction modulo $k$ of the class of
$(H,\omega)/2\pi$ in the degree $3$ integral relative cohomology of
$G$ modulo $Q$.  On the other hand, a $\U(1)$ gauge field on the
D-submanifold would give rise to a class in the degree $2$ integral
cohomology of $Q$, namely the first Chern class of the corresponding
line bundle.  These two cohomology groups are related by a long exact
sequence in cohomology and studying this sequence will reveal that the
integral cohomology class represented by $H/2\pi$ is the obstruction
to associate a line bundle on $Q$ with the relative class
$[(H,\omega)]/2\pi$.  In other words, when $H$ is topologically
nontrivial there is strictly speaking no line bundle on the
D-submanifold; but rather, as we will see, an equivalence class of
line bundles corresponding to the reduction modulo $k$ of their first
Chern classes, in such a way that their fluxes also compute the
D0-charge.  This has a natural interpretation in terms of gerbes.

This paper is organised as follows.  In Section~\ref{sec:DBWZW} we
briefly review the geometry of the WZW model with and without
boundary, paying particular attention to the consistency conditions
which must be met for the well-definedness of the quantum theory.  The
appropriate mathematical framework for discussing these conditions is
the relative cohomology of the Lie group modulo the D-submanifold.  In
Section~\ref{sec:Flux} we study the relation between the D0-charge
defined in \cite{Q0} and flux quantisation.  We will find that despite
the absence in general of a line bundle on the D-submanifold, the
modularity of the D0-charge is such that the flux also computes it.
This modularity has a natural interpretation in terms of gerbes.
Finally in Section~\ref{sec:conc} we summarise the main points of the
paper.

\section*{Acknowledgements}

It is a pleasure to thank Costas Bachas, Michael Douglas, Krzysztof
Gaw\c{e}dzki, Maximilian Kreuzer, Donald Marolf, Andrew Ranicki,
Andreas Recknagel, Elmer Rees, Christoph Schweigert, Michael Singer
and Arkady Tseytlin for correspondence.  In addition JMF would like to
thank the Spinoza Institute and particularly Bernard~de~Wit for the
kind hospitality while this paper was being written.  This work was
finished while on a visit to CERN.

\section{D-branes in the boundary WZW model}
\label{sec:DBWZW}

In this section we describe the boundary WZW model; that is, the WZW
model associated to a worldsheet with boundary \cite{KlS,Gaw}.  It is
instructive to compare this with the ``standard'' WZW model, so we
recall this briefly first.  Although these results are not new, we
believe it is useful to collect them here in a hopefully simpler form
than has been done until now.

\subsection{The WZW model}
\label{sec:WZW}

Let $\Sigma$ be a compact Riemann surface (without boundary) and let
$G$ be a Lie group admitting a bi-invariant metric.  In order to
circumvent special cases associated with abelian groups or with
noncompact groups, we will assume that $G$ is compact semisimple;
although the theory is of course more general.  The WZW model is the
theory of maps $g:\Sigma \to G$ defined by the following action:
\begin{equation}
  \label{eq:WZW}
  I = \int_\Sigma \left<g^{-1}\partial g,g^{-1}\Bar\partial g\right> +
  \int_M H~,
\end{equation}
where $M$ is a 3-dimensional submanifold of $G$ with boundary
$\partial M = g(\Sigma)$, and $H = \frac16
\left<\theta,[\theta,\theta]\right>$, where $\theta$ is the
left-invariant Maurer--Cartan $1$-form on $G$ and $\left<-,-\right>$
is an invariant metric on the Lie algebra of $G$.

There is an obstruction to the existence of $M$, which is measured by
the homology class of $g(\Sigma)$ in $H_2(G)$.  Demanding that this
class vanish for all $g(\Sigma)$ is equivalent to demanding that
$H_2(G)$ vanishes.  This is the case, for instance, for $G$ a compact
semisimple Lie group.\footnote{For compact Lie groups where $H=0$ or
  noncompact groups where $H$ is exact, one can work with the
  $B$-field directly, without having to introduce the submanifold $M$
  and hence avoiding any obstructions.  For these theories, it is the
  $B$-field that is part of the data and one does not impose that the
  theory be independent of this choice.}

Even when the obstruction is overcome and such an $M$ found, the
action depends on the choice of $M$; although because the $3$-form $H$
is closed, the equations of motion do not.  Indeed if $M'$ is another
$3$-dimensional submanifold of $G$ with boundary $g(\Sigma)$, then
$M-M'$ is a $3$-cycle and as observed originally in \cite{WW}, the
path integral is also independent on the choice of extension provided
that
\begin{equation*}
  \int_{M-M'} H \in 2\pi\ZZ~.
\end{equation*}
Demanding that this be the case for all $3$-cycles $M-M'$, is
equivalent to the cohomology class $[H]/2\pi \in H^3(G;\RR)$ being
integral.  The definition of $H$ (as well as the kinetic term)
involves a choice of bi-invariant metric.  For $G$ a simple Lie group
there is a unique conformal class of bi-invariant metrics, and it is
always possible to choose a metric in this class for which this
integrality condition is satisfied.

\subsection{The boundary WZW model}
\label{sec:BWZW}

Suppose now that $\Sigma$ is a compact Riemann surface with nonempty
boundary $\partial\Sigma$.  The boundary is homeomorphic to a disjoint
union of circles.  We will assume for simplicity of exposition that
the boundary is connected, so that it consists of only one circle.
The extension to the general case does not represent any added
difficulties nor does it reveal any extra structure at this level.

\begin{figure}[h!]
\setlength{\unitlength}{1pt}
\centering
\begin{picture}(200,150)(0,0)
\put(0,0){\epsfig{file=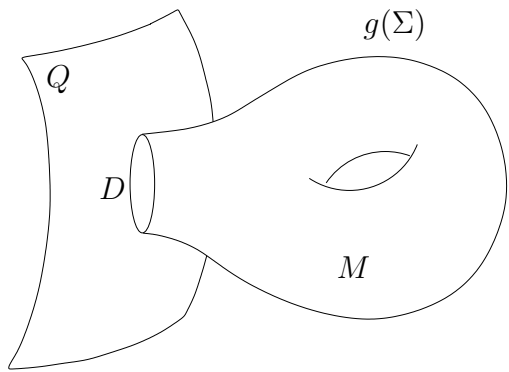,width=200pt}}
\put(20,110){\makebox(0,0)[bl]{$Q$}}
\put(140,130){\makebox(0,0)[bl]{$g(\Sigma)$}}
\put(40,70){\makebox(0,0)[bl]{$D$}}
\put(130,40){\makebox(0,0)[bl]{$M$}}
\end{picture}
\caption{The relation $\partial M = g(\Sigma) + D$.  In the figure $M$ is
  the solid object whose boundary is $g(\Sigma) + D$, where $D$ is
  contained in the D-submanifold $Q$.}
\label{fig:BWZW}
\end{figure}

Dirichlet boundary conditions in the WZW model are described
\cite{KlS} by a submanifold $\iota: Q \into G$ of the Lie group
\emph{and} a $2$-form $\omega$ on $Q$ such that $\iota^*H = d\omega$.
The submanifold $Q$ is called a D-submanifold.  The WZW model
corresponding to this boundary condition is the theory of maps
$g:\Sigma \to G$ sending the boundary $\partial\Sigma$ to $Q$, which
is governed by the following action
\begin{equation}
  \label{eq:BWZW}
  I = \int_\Sigma \left<g^{-1}\partial g,g^{-1}\Bar\partial g\right> +
  \int_M H - \int_D \omega~,
\end{equation}
where $M$ is a $3$-dimensional submanifold of $G$ with boundary
$\partial M = g(\Sigma) + D$ where $D$ is a $2$-dimensional
submanifold of $Q$.  Applying the boundary again, we see that
$g(\Sigma)$ and $D$ have the same boundary with opposite orientations.
Therefore we can think of $\partial M$ as the manifold
$g(\Sigma)\cap_\partial D$ obtained by gluing the worldsheet
$g(\Sigma)$ and $D$ along their common boundary, as in
Figure~\ref{fig:BWZW}.

As in the case of the WZW model, there is a homological obstruction to
the existence of $M$.  This time the relevant homology theory is the
relative homology of $G$ modulo $Q$.  Since the boundary of the
worldsheet $g(\Sigma)$ of the string lies in $Q$, $g(\Sigma)$ is a
relative $2$-cycle and defines a homology class in $H_2(G,Q)$.  The
existence of $M$ and $D \subset Q$ such that $\partial M = g(\Sigma) +
D$ simply says that $g(\Sigma)$ ia a boundary modulo $Q$, whence its
relative homology class is zero.  Demanding that this be true for all
$g(\Sigma)$ is equivalent to demanding that $H_2(G,Q)$ vanish.

Even when the existence of $M$ (and hence $D$) is unobstructed, such
$M$ is generally not unique and the action will depend on the choice
of $M$. As in the WZW model, the equations of motion do not depend on
the choice of $M$. (In fact, they do not depend on $\omega$ either,
only the boundary conditions do.)  A condition which guarantees the
independence of the path integral on the choice of $M$ can again be
captured cohomologically---this time in the relative cohomology of the 
pair $(G,Q)$ \cite{KlS,Gaw}.

\begin{figure}[h!]
\setlength{\unitlength}{1pt}
\centering
\begin{picture}(320,300)(0,0)
\put(0,0){\epsfig{file=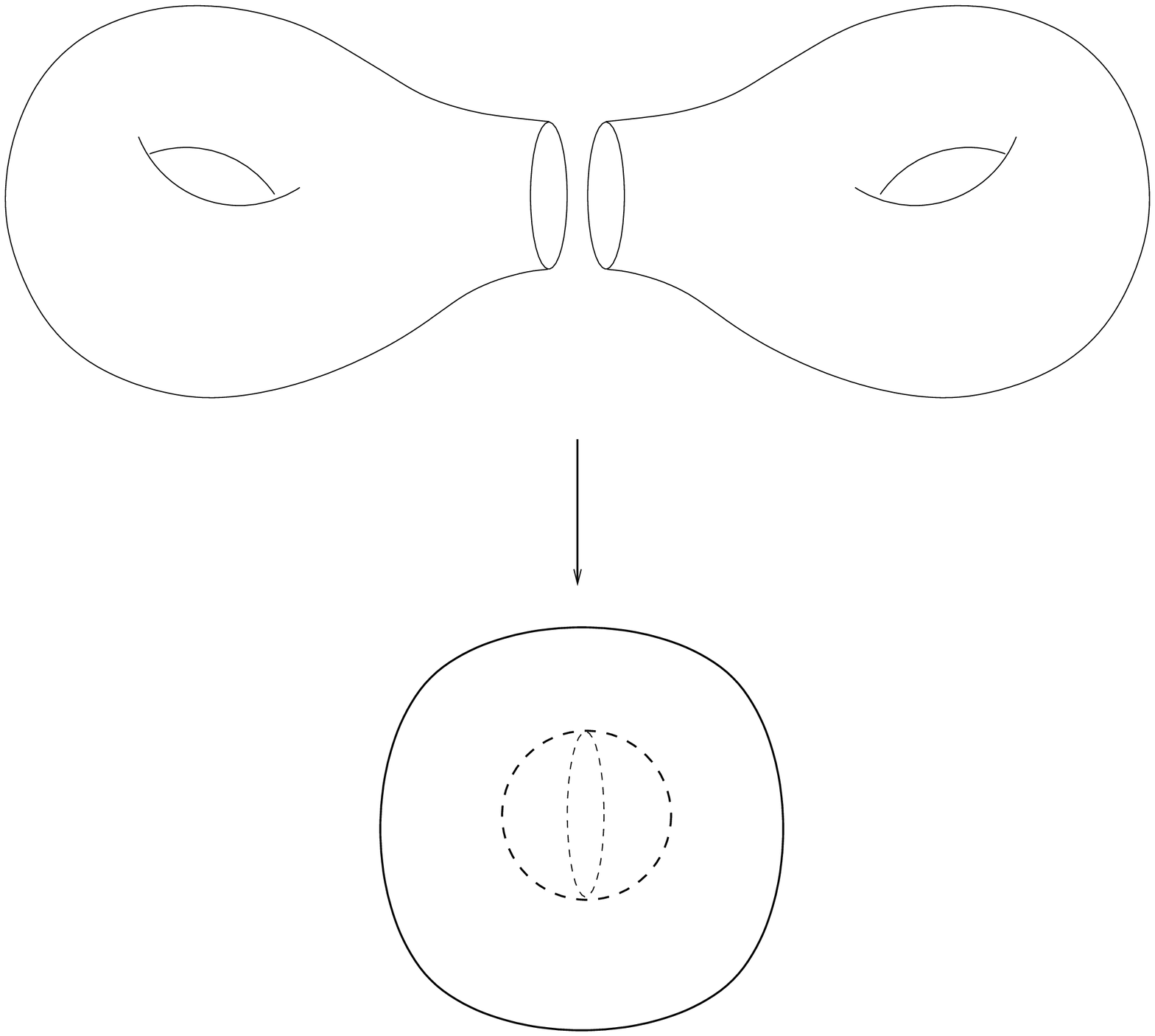,width=310pt}}
\put(170,220){\makebox(0,0)[bl]{$D_2$}}
\put(125,220){\makebox(0,0)[bl]{$D_1$}}
\put(60,190){\makebox(0,0)[bl]{$M_1$}}
\put(250,190){\makebox(0,0)[bl]{$M_2$}}
\put(100,270){\makebox(0,0)[bl]{$g(\Sigma)$}}
\put(200,270){\makebox(0,0)[bl]{$g(\Sigma)$}}
\put(165,60){\makebox(0,0)[bl]{$S$}}
\put(120,80){\makebox(0,0)[bl]{$Z$}}
\end{picture}
\caption{Gluing $M_1$ and $M_2$ along $g(\Sigma)$ to obtain the
  relative cycle $Z$, and gluing $D_1$ and $D_2$ along
  $g(\partial\Sigma)$ to obtain its boundary $\partial Z= S$.}
\label{fig:gluing}
\end{figure}

To see this let $M_1$ and $M_2$ be two $3$-dimensional submanifolds of
$G$ such that $\partial M_1 = g(\Sigma) + D_1$ and $\partial M_2 =
g(\Sigma) + D_2$ for some $D_1$ and $D_2$ in $Q$.  Then the difference
in the Wess--Zumino term is
\begin{equation*}
  \left(\int_{M_1} H - \int_{D_1} \omega\right) - \left( \int_{M_2} H
  - \int_{D_2}   \omega\right) = \int_{M_1-M_2} H - \int_{D_1-D_2}
    \omega~.
\end{equation*}
Let $Z = M_1 - M_2$ and $S=D_1 - D_2$.  Notice that
\begin{equation*}
  \partial Z = \partial M_1 - \partial M_2 = \Sigma + D_1 - \left( \Sigma + D_2\right) = 
  D_1 - D_2 = S~.
\end{equation*}
Since $S\subset Q$, $Z$ is a cycle modulo $Q$, whence it defines a
class in the degree $3$ relative homology of $G$ modulo $Q$.  We can
picture $Z$ as obtained by gluing $M_1$ and $M_2$ along $g(\Sigma)$
and $S$ as gluing $D_1$ and $D_2$ along their boundary
$g(\partial\Sigma)$, as shown in Figure~\ref{fig:gluing}.

The path integral is independent of the choice of $Z$ (and hence $S$)
provided that the quantity
\begin{equation}
  \label{eq:GWA}
  C = \frac{1}{2\pi} \left(\int_{Z} H - \int_{S} \omega \right)
\end{equation}
is integral.  As we now explain, this is simply the pairing between
the relative homology and the relative cohomology of $G$ modulo $Q$
(see, e.g., \cite{KarLer,BottTu}.

The real relative cohomology of $G$ modulo $Q$ can be computed from
the relative de~Rham complex.  A relative $p$-form in this complex
consists of a pair $(\alpha,\beta)$ where $\alpha$ is a $p$-form on
$G$ and $\beta$ is a ($p-1$)-form on $Q$.  The differential is defined 
by $d(\alpha,\beta) = (d\alpha, \iota^* \alpha - d\beta)$ where
$\iota: Q\to G$ is the embedding and $\iota^*$ is the pullback on
forms.  It obeys $d^2=0$ and the resulting cohomology is the relative
de~Rham cohomology $H^*(G,Q;\RR)$ of $G$ modulo $Q$.  Notice that a
relative form $(\alpha,\beta)$ is closed if $\alpha$ is closed in $G$
and exact when restricted to $Q$: $\iota^* \alpha = d\beta$.

As in the de~Rham complex, integration provides the pairing between
cycles (more generally, currents) and closed forms.  Given a relative
$p$-cycle $N$ and a closed relative $p$-form $(\alpha,\beta)$ the
expression
\begin{equation*}
  \int_N \alpha - \int_{\partial N} \beta
\end{equation*}
gives a pairing between the relative homology and the relative de~Rham 
cohomology of $G$ modulo $Q$.

Therefore we see that the quantity $C$ in equation \eqref{eq:GWA} is
precisely the result of the pairing between the relative $3$-cycle $Z$
and the relative $3$-form $(H,\omega)/2\pi$.  Demanding that $C$ be
integral for all relative cycles $Z$ is simply the requirement that
the relative cohomology class $[(H,\omega)]/2\pi \in H^3(G,Q;\RR)$ be
integral, which is precisely the cancellation of the global worldsheet
anomaly \cite{FW}.  Table~\ref{tab:topology} contrasts the topological
conditions for the existence of the quantum WZW models with and
without boundary.\footnote{This table possibly suggests that the
  boundary WZW model should be renamed the \emph{relative} WZW model.
  Alas we have not been able to reach an agreement on this point.}

\begin{table}[h!]
\centering
\setlength{\extrarowheight}{3pt}
\begin{tabular}{|c|>{$}c<{$}|>{$}c<{$}|}\hline 
Model & \text{Obstruction} & \text{Well-definedness}\\
\hline\hline
WZW & H_2(G)=0 & [H]/2\pi \in H^3(G;\ZZ)\\
BWZW & H_2(G,Q)=0 & [(H,\omega)]/2\pi \in H^3(G,Q;\ZZ)\\
\hline
\end{tabular}
\vspace{8pt}
\caption{Topological conditions for the existence of the quantum WZW
model with and without boundary.}
\label{tab:topology}
\end{table}

In a theory of strings propagating in a Lie group, we want both closed
strings propagating in the bulk and open strings with ends in the
D-branes.  Therefore consistency of the theory requires that both the
WZW model and the boundary WZW model should be well-defined.  This
means that both $H_2(G)$ and $H_2(G,Q)$ should vanish; and that both
$[H]/2\pi \in H^3(G;\RR)$ and $[(H,\omega)]/2\pi \in H^3(G,Q;\RR)$
should be integral classes.

\subsection{D-branes in WZW models}
\label{sec:DWZW}

An interesting class of D-branes which are by now well-understood are
those where the D-submanifold $Q$ corresponds to a (possibly twisted,
shifted) conjugacy class.  These D-branes are special in that they
preserve not just conformal invariance but also (one half of) the
infinite-dimensional symmetry current algebra of the WZW model.  They
are described in terms of the following gluing conditions:
\begin{equation}
  \label{eq:gc}
  J = R\Bar J~,
\end{equation}
where $J = -\partial g g^{-1}$, $\Bar J = g^{-1}\Bar\partial g$ and
$R$ is a metric-preserving automorphism of the Lie algebra of $G$.
This type of gluing conditions describe \cite{AS,SAdS3,FFFS,SDnotes}
D-branes whose worldvolumes lie on twisted conjugacy classes
\begin{equation*}
  C_r(g_0) := \left\{ r(g) g_0 g^{-1} \mid g \in G\right\}~,
\end{equation*}
where $r:G \to G$ is the metric-preserving automorphism of $G$ which
integrates $R$.

Two metric-preserving automorphisms $r$ and $r'$ which are related by
an inner automorphism, yield twisted conjugacy classes $C_r(g_0)$ and
$C_{r'}(g_0)$ which are simply shifted relative to each other.  Hence
in this sense these types of D-submanifolds are classified \cite{FSNW}
by the group $\Out_o(G)$ of metric-preserving outer automorphisms of
$G$, which is defined as the quotient $\Aut_o(G)/\Inn_o(G)$ of the
group of metric-preserving automorphisms by the invariant subgroup of
inner automorphisms.

As shown in \cite{AS,Gaw,Q0} there is a natural $2$-form $\omega$ on
each twisted conjugacy class $C_r(g_0)$ such that $d\omega$ agrees
with the restriction of the $3$-form $H$ to $C_r(g_0)$.  This $2$-form
is obtained by demanding that the boundary conditions associated to
the gluing conditions \eqref{eq:gc} and the ones coming from the sigma
model description of the WZW model coincide.

\section{Flux quantisation and D0-charge}
\label{sec:Flux}

The issue of flux quantisation and the definition of the D0-charge for
D-branes in Lie groups has generated a great deal of interest
recently.  The existence of a $\U(1)$ gauge field on the D-brane whose
flux is quantised was assumed in \cite{BDS,Ta}, in order to discuss
the stability and D0-charge of the D2-branes in $\SU(2)$.
Alternatively, it was shown in \cite{Q0} that both the stability and
the various D0-charges of this type of D-branes can be analysed
without having to rely on the existence of a $\U(1)$ gauge field on
the D-brane, by defining all the relevant quantities in terms of the
globally defined gauge invariant fields $H$ and $\omega$.  The central
concept in this approach is the global worldsheet anomaly, whose
vanishing explains both the discrete spectrum of stable D-branes and
the quantisation of a suitably defined D0-charge.  However one
question remains: Does the vanishing of the worldsheet anomaly
actually imply the quantisation of a $\U(1)$ gauge field flux on the
brane?

This question is symbolically encoded in the following identity:
\begin{equation}
  \label{eq:que}
  \int_Z H - \int_S \omega \stackrel{?}{=} \int_S F~;
\end{equation}
where the precise definition and geometrical nature of the field $F$
in the right-hand side have hitherto remained somewhat obscure.  The
purpose of this section is to shed some light on this hypothetical
equivalence.

Before going into any systematic analysis let us make a rather simple
but instructive remark.  The left-hand side of equation \eqref{eq:que}
depends not only on $S$ but also on $Z$, whereas the right-hand side
depends on $S$ and on $F$.  Given an $S$ and a $Z$, it is possible to
find \emph{an} $F$ such that the identity in \eqref{eq:que} holds; but
this $F$ will depend on $Z$.  The existence of one $F$ for which this
identity holds irrespective of $Z$, means that the $Z$-dependence of
the left-hand side is only apparent and this imposes a condition on
$H$.  To see this let $Z'$ be another submanifold with boundary $S$.
The identity in \eqref{eq:que} would result in the following condition
on $H$
\begin{equation*}
  \int_{Z-Z'} H = 0~,
\end{equation*}
where $Z-Z'$ is a $3$-cycle, and this would imply that $H$ is exact.
This suggests that when $H$ is not exact, the identity in
\eqref{eq:que} will not hold for a fixed $F$.

\subsection{A line bundle on the D-submanifold}
\label{sec:linebundles}

Clearly, the strong version of the statement encoded in \eqref{eq:que}
would be to have a line bundle on the D-submanifold whose Chern class
is equal to the relative class defined by the fields $H$ and $\omega$.
As we now explain, there are conditions under which the relative class
represented by $(H,\omega)/2\pi$ is equivalent to an integral class in
$H^2(Q)$; that is, to the Chern class of a line bundle on the
D-submanifold $Q$.  In this case, $C$ in \eqref{eq:GWA} can be
understood as the flux of $F/2\pi$, where $F$ is the curvature on the
line bundle.

Indeed, suppose that $H=dB$ is exact.  This does \emph{not} mean that
the relative cocycle $(H,\omega)$ is a coboundary.  Indeed, the
cocycle condition simply says that $\iota^*H = \iota^* dB = d\omega$
on $Q$, whence the $2$-form $F= \iota^*B - \omega$ defined on $Q$ is
closed and defines a class in $H^2(Q)$.  If $W$ is any $3$-dimensional
submanifold of $G$ whose boundary $\partial W$ is contained in $Q$,
then using Stokes we have that
\begin{equation*}
  \int_W dB - \int_{\partial W} \omega = \int_{\partial W} F~,
\end{equation*}
and hence integrality of $[(dB,\omega)]/2\pi\in H^3(G,Q;\RR)$ is
precisely the integrality of $[F]/2\pi\in H^2(Q;\RR)$.
\footnote{Strictly speaking, the above equality only proves that
  $[F]/2\pi$ is an integer on those $2$-cycles in $Q$ which bound in
  $G$.  Since we are assuming, for consistency of the WZW model that
  $H_2(G)=0$ this is true for all $2$-cycles on $Q$.}  This is
equivalent to the existence of a line bundle on $Q$ whose curvature is
$F$ and whose first Chern class is $[F]/2\pi$.

Furthermore, the converse is also true and a relative class
$[(H,\omega)]$ in $H^3(G,Q)$ comes from a class in $H^2(Q)$
\emph{only} if the integral class represented by $H/2\pi$ is zero.
\footnote{Notice that this is stronger than the fact that $H$ should
  be exact, since this only means that the class $[H]/2\pi$ is
  torsion.  Of course, if $G$ is simply connected, $G$ is homotopy
  equivalent to a product of odd spheres and hence $H^3(G)$ has no
  torsion.}  This is a consequence of the exact cohomology sequence
(see, e.g., \cite{BottTu})
\begin{equation}
  \label{eq:ECS}
  \cdots \to H^2(G) \xrightarrow{\iota^*} H^2(Q) \xrightarrow{} 
  H^3(G,Q) \xrightarrow{} H^3(G) \to \cdots
\end{equation}

Indeed, consider a relative class $[(H,\omega)]/2\pi$ in $H^3(G,Q)$.
By exactness at $H^3(G,Q)$, such a class comes from a class in
$H^2(Q)$---that is, is in the image of $H^2(Q) \to H^3(G,Q)$---if and
only if it is the kernel of $H^3(G,Q)\to H^3(G)$; in other words if
the class $[H]/2\pi$ in $H^3(G)$ vanishes.

In other words, we can understand the integral cohomology class in
$H^3(G)$ represented by $H/2\pi$ as the obstruction to defining a line
bundle $L$ on $Q$ whose first Chern class obeys $c_1(L) =
[(H,\omega)]/2\pi$.  From this it follows in particular that, for a
compact semisimple Lie group, there is no line bundle on a D-brane
described by a conjugacy class.

\subsection{Local considerations}
\label{sec:local}

It is important to remark that even if $H$ is not exact, its
restriction to $M$ is.  This follows from the fact that a top form on
a manifold with nonempty boundary is exact, and the fact that $H$ is a
$3$-form and $M$ is a $3$-manifold with nonempty boundary $g(\Sigma)
\cup_\partial D$.  Therefore there exists a $2$-form $B$ on $M$ such
that $H = dB$.  This means that there is a closed $2$-form $F = B -
\omega$ on $D$ such that the equation \eqref{eq:que} is satisfied.
The triviality of $H^3(M)$ is true also in integral cohomology,
therefore there is a line bundle on $D$ whose curvature is $F$, and
the integrality of $C$ is the quantisation of the flux of $F$.  It is
important to emphasise that this line bundle is defined on $D$ and
\emph{not} on all of $Q$.  In other words, whereas $\omega$ is
intrinsic to $Q$, $B$ is intrinsic to $M$ and hence $F$ is only
defined where both of these quantities make sense; that is, on $D=M
\cap Q$.  This seems to suggest a physical picture of the $\U(1)$
field on the brane being due to the ends of the open strings ending on
the D-brane and hence existing locally near them.

It might be illuminating to compare this with the case of the standard
WZW model.  Although $H$ is in general not an exact form on the group
$G$, it again becomes exact when restricted to $M$, the
$3$-dimensional submanifold of $G$ whose boundary is $\partial M =
g(\Sigma)$.  This means that there is a $2$-form $B$ in $M$ which
satisfies $H = dB$ there.  In terms of this form, the action can be
written in a manifestly local way \cite{WW}, since
\begin{equation}
  \label{eq:WZ} 
  \int_M H = \int_M dB = \int_{g(\Sigma)} B = \int_\Sigma g^*B~.
\end{equation}
Notice that $B$ is not generally the restriction to $M$ of a $2$-form
defined on the whole group, as this would require $H$ to be exact.

\subsection{D0-charge}
\label{sec:Q0}

Several definitions of the D0-charge have been proposed recently. The
picture that begins to emerge is that one can in fact introduce
several D0-charges \cite{Marolf1}, distinguished not only by their
form but also by their specific properties (e.g., gauge invariance,
quantisation).  In the sigma model framework one can define at least
two types of charges: the so-called brane source charge introduced in
\cite{BDS}, which is gauge invariant but not quantised, and a Page
charge \cite{AMM,Q0} which is gauge invariant (at least
infinitesimally) and quantised.\footnote{There is a third type of
  charge, the Maxwell charge, which is computed in \cite{Ta} and which
  agrees with the Page charge in the case of D2-branes in the WZW
  model.}

The brane source charge can be written as the integral of the two-form
$\omega$ on a 2-cycle of the D-submanifold.  It is in general not
conserved because $\omega$ is not closed.  Nevertheless its
nonconservation is to a large extent under control since
$d\omega = H$.  In fact this very relation suggests us a way of
modifying the brane source charge in order to obtain a conserved
quantity, which turns out to be nothing but $C$.  This procedure is
reminiscent of the way one constructs the Page charge in supergravity.
Notice however that, as we pointed out before, $C$ depends not only on 
$H$, $\omega$ and $S\subset Q$, which describe the given D-brane
configuration, but it depends also on $Z$, the $3$-submanifold of $G$
with boundary $S$.

An alternative way of motivating the definition of the (Page)
D0-charge in terms of the quantity $C$ in \eqref{eq:GWA} is the
following.  We have seen that $C$ agrees, when $H$ is exact, with the
normalised flux of a $\U(1)$ gauge field $F$ on $Q$.  It is therefore
natural in this case to identify $C$ with the D0-charge of a region $S
\subset Q$: $\int_S F/2\pi$.  This point of view led in \cite{Q0} to
the proposal that the expression for $C$ in equation \eqref{eq:GWA}
should be understood as a covariantisation of $\int_S F/2\pi$ in the
case where $H$ is not exact.  However, as noted above, this expression
depends on $Z$, the $3$-submanifold of $G$ with boundary $S$.

Physically the D0-charge of the region $S$ should not depend on $Z$.
However, as we saw above, the difference in the charges computed by
$Z$ and $Z'$ is the integral of $H/2\pi$ on the $3$-cycle $Z-Z'$ of
$G$.  For $G$ a compact simple Lie group, $H^3(G) \cong \ZZ$.
Therefore if the $3$-cycle $Z-Z'$ is mapped to $n$ under this
isomorphism, one has
\begin{equation*}
  \frac{1}{2\pi} \int_{Z-Z'} H = n\, k~,
\end{equation*}
which is always a multiple of the level $k$.  Therefore, if we insist
in the $Z$-independence of the D0-charge, we must define it to be
given by $C$ (with an appropriate normalisation) and reduce it modulo
$k$.  In other words, the D0-charge is the image of
$[(H,\omega)]/2\pi$ under the natural map $H^3(G,Q;\ZZ) \to
H^3(G,Q;\ZZ_k)$ induced by reduction modulo $k$.

In \cite{Q0}, for the special case of $G=\SU(2)$ this modularity was
found to be consistent with the fact that the D0-branes sitting at the
points in the center of the group should carry the same charge.  This
modularity of the D0-charge was also observed in \cite{AS2} in the
context of the supersymmetric WZW model, where the level is shifted to
$k+2$ for $G=\SU(2)$.

For a compact semisimple Lie group $G=\prod_{i=1}^n G_i$, with $G_i$
simple and with level $k_i$, the same argument implies that the
D0-charge should be defined modulo the greatest common divisor $k =
\gcd(k_1,\dots,k_n)$ of the levels.  Notice that this has the property
that if two of the levels are coprime, then the charge is always zero.

Defining the charge in this way also has the virtue that the
hypothetical equality in \eqref{eq:que} becomes an honest equality
modulo $k$.  In this way, one could still obtain the D0-charge from
the flux of a (locally defined) $\U(1)$ gauge field, since the change
in the flux on overlaps is a multiple of $k$ and hence not seen by the
charge.

\subsection{Gerbe interpretation}
\label{sec:gerbe}

The fact that there is no canonical line bundle on the D-submanifold
should not come as a surprise.  At the heart of the WZW model is an
integral class in $H^3(G)$, which should alert us to the existence of
an underlying gerbe (see, for example, \cite{Brylinksi}).  Gerbes are
the third member in an infinite sequence of objects, whose first two
members are smooth circle-valued functions and line bundles,
respectively.  They are an attempt to geometrise integral classes in
$H^3$ in roughly the same way that line bundles geometrise integral
classes in $H^2$.  There are several different descriptions of gerbes.

One such description is in terms of line bundles on the loop group of
$G$.  By transgression (see, e.g., \cite{PS}) integral classes in
$H^3(G)$ are in bijective correspondence with integral classes in
$H^2(LG)$, where $LG$ is the loop group of $G$.  Therefore one can
understand a gerbe in $G$ as a line bundle on $LG$.  This description
is useful in discussing the functional integral approach to the WZW
model \cite{Gaw}.  This line bundle can be trivialised locally on the
subspace $LQ \subset LG$ corresponding to loops in the conjugacy
class; but trivialising a gerbe does not give rise to a line bundle on
$Q$.

To see this one needs a different description of a gerbe, in terms of
(locally defined) line bundles on $G$ \cite{Hitchin}.  Suppose that
$\eU = \left\{U_\alpha\right\}$ is an open cover for $G$ such that the
restriction of the $3$-form $H$ to $U_\alpha$ is exact:
$H\bigr|_{U_\alpha} = d B_{\alpha}$.  The gerbe on $G$ characterised
topologically by the integral class $[H]/2\pi$ in $H^3(G)$ can be
defined by specifying a line bundle $L_{\alpha\beta}$ on each double
intersection $U_\alpha\cap U_\beta$, an isomorphism $L_{\alpha\beta}
\cong L_{\beta\alpha}^{-1}$, and some further conditions in triple and
fourfold intersections which will not concern us here.  These line
bundles come equipped with a natural connection whose curvature is
given by $f_{\alpha\beta} = B_\alpha - B_\beta$.

Now let $\eV = \left\{ V_\alpha = U_\alpha \cap Q\right\}$ be an open
cover for $Q$ induced from the one for $G$.  On $V_\alpha$ define
$F_\alpha = B_\alpha - \omega$, since the $2$-form $\omega$ is
globally defined on $Q$.  Clearly $d F_\alpha = 0$, and it is this
that is interpreted as the curvature of the hypothetical line bundle
on $Q$.  However the curvature is not globally defined on $Q$: on
$V_\alpha \cap V_\beta$, $F_\alpha - F_\beta = f_{\alpha\beta}$.  If
the class of $H$ is nontrivial, neither is $f_{\alpha\beta}$ and
$F_\alpha$ and $F_\beta$ are therefore curvatures on topologically
different line bundles.  This means that the line bundles of which the
$\{F_\alpha\}$ are the curvatures do not patch up to a global line
bundle.

Let us illustrate this for $G=\SU(2)\cong S^3$, where $Q\cong S^2$ is
a spherical conjugacy class.  It separates $G$ into two hemispheres
$G_\pm$ with common boundary $Q$.  Let $U_\pm$ be open sets defined as
the complement of chosen points in the interior of $G_\mp$
respectively.  For example, if we think of $Q$ as a ``parallel'', we
can take $U_+$ to be the complement of the south pole and $U_-$ to be
the complement of the north pole. The intersection is $U_+ \cap U_-
\cong Q \times \RR$.  On $U_\pm$ we have $H=d B_\pm$, and on $U_+ \cap
U_-$ we have $B_- - B_+ = f_{+-}$, which is the curvature of a line
bundle defined on $Q \times \RR$.  The restriction of this line bundle
to $Q$ is topologically nontrivial provided that $H$ is as well.

\begin{figure}[h!]
\setlength{\unitlength}{1pt}
\centering
\begin{picture}(200,150)(0,0)
\put(0,0){\epsfig{file=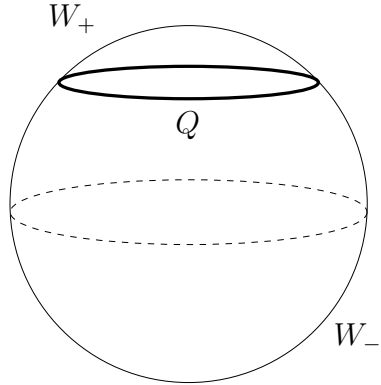,width=150pt}}
\put(70,100){\makebox(0,0)[bl]{$Q$}}
\put(40,140){\makebox(0,0)[br]{$W_+$}}
\put(130,30){\makebox(0,0)[tl]{$W_-$}}
\end{picture}
\caption{Two ways of computing the D0-charge of a spherical D2-brane
$Q$ in $\text{SU}(2)$.}
\label{fig:hemispheres}
\end{figure}

To see this, let $W_\pm$ be closed sets contained in $U_\pm$ covering
$G$ and such that $W_+ \cap W_- = Q$ as shown in
Figure~\ref{fig:hemispheres}. We then integrate to find:
\begin{equation*}
  \int_G H = \int_{W_+} H + \int_{W_-} H = \int_Q B_+ - \int_Q B_- =
  \int_Q f_{+-}~.
\end{equation*}
Since $\int_G H = 2\pi k$, it follows that the first Chern class of
the line bundle whose curvature is $f_{+-}$ is $k$ times the generator
of $H^2(Q)\cong \ZZ$.  This provides another way to understand why for
the $\SU(2)$ D2-branes, the flux is only defined modulo $k$.

\section{Conclusions}
\label{sec:conc}

In this paper we have re-examined the issues of D0-charge and flux
quantisation in the context of D-branes in WZW models.  We have
started by studying the topological conditions necessary for the
existence of a consistent theory of open and closed strings
propagating on a Lie group.  The topological conditions are phrased
naturally in terms of the (relative) cohomology of the Lie group
modulo the D-submanifold on which the D-branes wrap.
Table~\ref{tab:topology} summarises these conditions. The upshot is
that that both $H^2(G)$ and $H^2(G,Q)$ must vanish and that not just
must the NS $3$-form $H/2\pi$ represent a class in $H^3(G;\ZZ)$, but
also the pair $(H,\omega)/2\pi$ must represent a class in the relative 
cohomology $H^3(G,Q;\ZZ)$.

For $G$ be a compact simple Lie group and $k$ the level of the
corresponding WZW model, the D0-charge is the class in
$H^3(G,Q;\ZZ_k)$ induced by the reduction modulo $k$ of the relative
class represented by $(H,\omega)/2\pi$ in $H^3(G,Q;\ZZ)$.  We then
showed that the class in $H^3(G;\ZZ)$ represented by $H/2\pi$ is the
obstruction to the existence of a line bundle on $Q$ whose first Chern
class induces the relative class $[(H,\omega)]/2\pi$.  Hence if $H$ is
not cohomologically trivial, there is no canonical line bundle on $Q$,
and hence no $\U(1)$ gauge field whose flux computes the D0-charge.
Instead we have a family of (locally defined) line bundles on $Q$,
interpreted as a trivialisation on $Q$ of the gerbe whose
characteristic class is $[H]/2\pi$, whose Chern classes are equivalent
modulo $k$ and such that they agree with the D0-charge.  We conclude
therefore that for nontrivial $H$, there is no line bundle on the
D-submanifold and hence no associated $\U(1)$ gauge theory on all of
$Q$.  Nevertheless there are locally-defined line bundles and gauge
fields whose fluxes are quantised (and defined modulo $k$) in such a
way that they agree with the D0-charge.

This local picture of the gauge theory suggests a physical situation
in which the gauge fields are indeed generated locally where the
strings hit the D-brane, but in such a way that they patch up globally
only modulo $k$.  It would be nice to have independent confirmation of
the modularity of the D0-brane charge.

Although we have concentrated in the WZW model, where $G$ is a Lie
group, many of the results in this paper are valid in more general
situations where $G$ is a riemannian manifold with a nontrivial NS
$3$-form.  The question of the nature of the ``gauge field'' on the
D-submanifold in the presence of a nontrivial NS $3$-form has been
studied also in \cite{FW,Kap,BouwMat} and our work gives a
complementary perspective on this issue.

%
%\bibliographystyle{utphys}
%\bibliography{AdS3,Geometry}

\providecommand{\href}[2]{#2}\begingroup\raggedright\endgroup

\end{document}